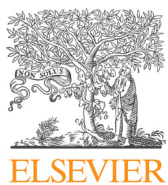
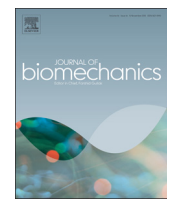

# A new strain energy function for the hyperelastic modelling of ligaments and tendons based on fascicle microstructure

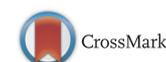

Tom Shearer *

*School of Mathematics, University of Manchester, Manchester M13 9PL, United Kingdom*## ARTICLE INFO

*Article history:*
Accepted 21 November 2014

*Keywords:*
Ligaments
Tendons
Strain energy function
Microstructure
Crimp angle distribution## ABSTRACT

A new strain energy function for the hyperelastic modelling of ligaments and tendons based on the geometrical arrangement of their fibrils is derived. The distribution of the crimp angles of the fibrils is used to determine the stress–strain response of a single fascicle, and this stress–strain response is used to determine the form of the strain energy function, the parameters of which can all potentially be directly measured via experiments – unlike those of commonly used strain energy functions such as the Holzapfel–Gasser–Ogden (HGO) model, whose parameters are phenomenological. We compare the new model with the HGO model and show that the new model gives a better match to existing stress–strain data for human patellar tendon than the HGO model, with the average relative error in matching this data when using the new model being 0.053 (compared with 0.57 when using the HGO model), and the average absolute error when using the new model being 0.12 MPa (compared with 0.31 MPa when using the HGO model).

© 2014 The Author. Published by Elsevier Ltd. This is an open access article under the CC BY license (http://creativecommons.org/licenses/by/3.0/).## 1. Introduction

Ligaments and tendons are fundamental structures in the musculoskeletal systems of vertebrates. Ligaments connect bone to bone to provide stability and allow joints to function correctly, whereas tendons connect bone to muscle to allow the transfer of forces generated by muscles to the skeleton. The wide variety of roles played by different ligaments and tendons requires them to have considerably different mechanical responses to applied forces, and their differing stress–strain behaviours have been well documented (Benedict et al., 1968; Tipton et al., 1986).

Ligaments and tendons consist of collagenous fibres organised in a hierarchical structure (Kastelic et al., 1978). Their main subunit is the fascicle, which consists of fibrils arranged in a crimped pattern (see Fig. 1). Further subunits in the hierarchy can be observed; however, the mechanics on these lengthscales will not be considered here. Instead, we shall focus on the effect of the geometrical arrangement of the fibrils within fascicles on the stress–strain properties of ligaments and tendons.

From a modelling perspective, ligaments and tendons can be categorised as fibre-reinforced biological soft tissues. A wide variety of models has been proposed to describe such tissues; however, to the author's knowledge, there has not yet been a successful attempt to develop a constitutive model within a non-linear elastic framework that includes the required anisotropy and characteristic stress–strain behaviour, which is non-linear with increasing stiffness for small strains (this region of the stress–strain curve is commonly termed the *toe region*) and subsequently linear, and, crucially, depends *only on directly measurable parameters*. Existing models are either phenomenological (Fung, 1967; Holzapfel et al., 2000), or lacking in the required material properties (such as the neo-Hookean model, which was developed for modelling rubber, but has still been used extensively in modelling biological soft tissues Miller, 2001, 2005), or both (Gou, 1970).

Early work on modelling biological tissue was carried out by Fung (1967). Fung showed that the stress in rabbit mesentery under uniaxial tension appears to increase exponentially as a function of increasing stretch. This exponential stress–strain relationship appears to approximate the behaviour of many biological soft tissues well, but only in a phenomenological sense and there is no microstructural basis for the choice of the exponential function. In 1970, Gou built upon Fung's work and proposed an isotropic strain energy function (SEF) for biological tissues that similarly gives an exponential stress–strain relationship in the case of uniaxial tension, but since this model is isotropic, it is not suitable for modelling anisotropic tissues such as ligaments and tendons.

With regard specifically to ligaments and tendons, various models were proposed over the following decades, as summarised in the review article by Woo et al. (1993). The models proposed involved infinitesimal elasticity (Frisen et al., 1969), finite elasticity (Hildebrant

* Tel.: +44 161 275 5810.
E-mail address: tom.shearer@manchester.ac.uk

http://dx.doi.org/10.1016/j.jbiomech.2014.11.0310021-9290/© 2014 The Author. Published by Elsevier Ltd. This is an open access article under the CC BY license (http://creativecommons.org/licenses/by/3.0/).

**Nomenclature**

| | |
|---|---|
| $W$ | strain energy function |
| $c, k_1, k_2$ | material parameters of Holzapfel–Gasser–Ogden model |
| $I_1, I_2$ | isotropic strain invariants |
| $I_4$ | anisotropic strain invariant |
| **B**, **C** | left/right Cauchy–Green tensor |
| **M**, **m** | direction of fascicles in undeformed/deformed configuration |
| **F** | deformation gradient tensor |
| $c_i$ | phenomenological material parameters |
| $T$ | $c_2(I_1-3)^2 + c_3(I_4-1)^2 + c_4(I_1-3)(I_4-1)$ |
| $\tilde{a}$ | fascicle radius |
| $\tilde{\rho}, \rho$ | dimensional/non-dimensional radial variable in fascicle |
| $\theta_p(\rho), \hat{\theta}_p(\rho)$ | fibril crimp angle distributions |
| $\theta_o, \theta_i$ | crimp angle of outermost/innermost fibrils |
| $\alpha, p$ | crimp angle distribution parameters |
| $\epsilon_p(\rho)$ | strain in fascicle as fibrils at radius $\rho$ become taut |
| $\epsilon^*$ | strain in fascicle as outer fibrils become taut |
| $b$ | crimp blunting factor |
| $\epsilon$ | given strain in fascicle |
| $R_p$ | radius within which all fibrils are taut for a given $\epsilon$ |
| $P_p$ | tensile loads experienced by fascicle |
| $\sigma_p(\rho), \epsilon_p^f(\rho)$ | stress/strain in fibrils at radius $\rho$ |
| $E^*$ | Hooke's law parameter utilised by Kastelic et al. (1980) |
| $\Delta\epsilon_p(\rho)$ | "elastic deformation" at radius $\rho$ |
| $E$ | Young's modulus of fibrils |
| $l_p(\rho), L$ | initial fibril/fascicle length |
| $\Delta l_p(\rho), \Delta L$ | fibril/fascicle extension |
| $\tau_p, \hat{\tau}_p$ | average tractions in the direction of the fascicle |
| $\lambda$ | stretch in the direction of the fascicle |
| $\beta$ | $2(1-\cos^3\theta_o)/(3\sin^2\theta_o)$ |
| **T**, **T**$^{HGO}$ | Cauchy stresses |
| $J$ | det **F** |
| $Q, Q^{HGO}$ | Lagrange multipliers |
| $\phi$ | fibre volume fraction |
| **T**$_f$, **t**$_f$ | component of stress/traction associated with fascicles |
| $\hat{\mathbf{m}}$ | unit vector in direction of **m** |
| $\gamma, \eta$ | constants of integration, defined in Eq. (51) |
| $\mu$ | ground state shear modulus of ligament/tendon matrix |
| $R, \Theta, Z$ | coordinate variables in undeformed configuration |
| $r, \theta, z$ | coordinate variables in deformed configuration |
| $A, a$ | undeformed/deformed radius of ligament/tendon |
| $B, b$ | undeformed/deformed length of ligament/tendon |
| $\zeta$ | stretch in longitudinal direction of ligament/tendon |
| **E**$_R$, **E**$_\Theta$, **E**$_Z$ | basis vectors in undeformed configuration |
| **e**$_r$, **e**$_\theta$, **e**$_z$ | basis vectors in deformed configuration |
| **n** | outer unit normal to curved surface of ligament or tendon |
| $S_{zz}, S_{zz}^{HGO}$ | longitudinal nominal stresses |
| $e$ | engineering strain |
| $\delta, \delta^{HGO}$ | relative errors |
| $\Delta, \Delta^{HGO}$ | absolute errors |

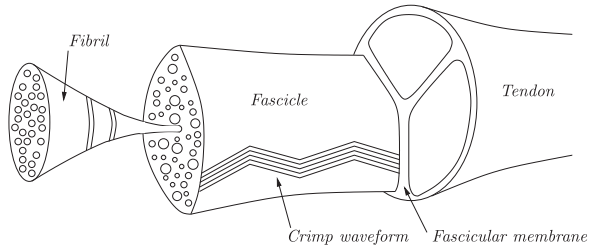

**Fig. 1.** Tendon hierarchy (adapted from Kastelic et al., 1978).

et al., 1969), quasi-linear viscoelasticity (Fung, 1968) and single integral finite strain viscoelasticity theory (Johnson et al., 1992). In particular, we note the work of Kastelic et al. (1980), in which a model was developed for the stress–strain response of a fascicle, taking into account the distribution of fibril crimp. It was shown that a radial variation in the crimp angle of a fascicle's fibrils could lead to a non-linear stress–strain relationship of the form typically observed in tension tests. Unfortunately, however, an error in the implementation of Hooke's law in that paper led to the derived relationship being incorrect, as we discuss further in Section 2.

Arguably the most influential model to be developed in the last 20 years for modelling biological tissues is the SEF proposed by Holzapfel et al. (2000), often referred to as the Holzapfel–Gasser–Ogden (HGO) model:

$$W = \frac{c}{2}(I_1 - 3) + \frac{k_1}{k_2}(e^{k_2(I_4-1)^2} - 1), \quad (1)$$

where $I_1$ and $I_4$ are strain invariants, defined by

$$I_1 = \mathrm{tr}\,\mathbf{C}, \quad \text{and} \quad I_4 = \mathbf{M} \cdot (\mathbf{CM}), \quad (2)$$

where $\mathbf{C} = \mathbf{F}^T \mathbf{F}$ is the right Cauchy–Green tensor, where **F** is the deformation gradient tensor (Ogden, 1997), and **M** is a unit vector pointing in the direction of the tissue's fibres before any deformation has taken place, $c$, $k_1$ and $k_2$ are material parameters, and the above expression is only valid when $I_4 \geq 1$ (when $I_4 < 1$, $W = c/2(I_1-3)$). This SEF was proposed as a constitutive model for arteries and is commonly used, along with its variants (Holzapfel and Ogden, 2010) to model a wide variety of biological materials. The advantages of this model are clear – it retains an elegant mathematical simplicity, whilst also providing the required anisotropy and "exponential-shaped" stress–strain curve common to many biological materials; however, as it is a phenomenological model, the parameters $c$, $k_1$ and $k_2$ cannot be directly linked to measurable quantities, and therefore the model has restricted predictive capabilities.

A large number of phenomenological, transversely isotropic, non-linear elastic models of biological soft tissues have been proposed. The following models were collated by Murphy (2013), where the parameters $c_i$, $i = 1,2,3,4,5,6,7$ are material parameters that can be chosen to match experimental data. Humphrey and Lin (1987) proposed this strain energy function for modelling passive cardiac tissue:

$$W = c_1(e^{c_2(I_1-3)} - 1) + c_3(e^{c_4(I_4^{1/2}-1)^2} - 1). \quad (3)$$

Humphrey et al. (1990) proposed the following for passive myocardium:

$$W = c_1(I_4^{1/2}-1)^2 + c_2(I_4^{1/2}-1)^3 + c_3(I_1-3) + c_4(I_4^{1/2}-1)(I_1-3) + c_5(I_1-3)^2. \quad (4)$$

Fung et al. (1993) proposed

$$W = c_1(e^T - T - 1) + c_5(I_1-3)^2 + c_6(I_4-1)^2 + c_7(I_1-3)(I_4-1), \quad (5)$$

where

$$T = c_2(I_1-3)^2 + c_3(I_4-1)^2 + c_4(I_1-3)(I_4-1), \quad (6)$$

to model canine thoracic aorta tissue, and Chui et al. (2007) used a similar function to model porcine liver tissue:

$$W = c_1 \log(1-T) + c_5(I_1-3)^2 + c_6(I_4-1)^2 + c_7(I_1-3)(I_4-1). \quad (7)$$

Taber (2004) used

$$W = c_1(I_1-3) + c_2(I_2-3) + c_4(I_4-1)^2, \quad I_2 = \frac{1}{2}(I_1^2 - \text{tr}(\mathbf{C}^2)), \quad (8)$$

to model cardiac papillary muscles and the embryonic heart, and the simpler function

$$W = c_1(I_1-3) + c_4(I_4-1)^2, \quad (9)$$

has been used by many authors, for example, Destrade et al. (2008), Ning et al. (2006), and Rohrle and Pullan (2007). Weiss et al. (1996) proposed a strain energy function whose anisotropic component was expressed in terms of derivatives with respect to the fascicle stretch $\lambda$:

$$\begin{aligned} \lambda W_\lambda &= 0, \quad \lambda < 1 \\ \lambda W_\lambda &= c_3(e^{c_4(\lambda-1)}-1), \quad 1 < \lambda < \lambda^* \\ \lambda W_\lambda &= c_5\lambda + c_6, \quad \lambda > \lambda^* \end{aligned} \quad (10)$$

where $W_\lambda$ represents the derivative of $W$ with respect to $\lambda$ and $\lambda^*$ is the critical stretch at which the non-linear region of the stress–strain curve ends. This strain energy function was then utilised by Peña et al. (2006) to investigate the effect of initial strains on the properties of biological tissues. All of the above strain energy functions have been shown to agree with experimental data reasonably well. The issue common to all of them, however, is that the material parameters $c_i$, $i=1,2,3,4,5,6,7$ have no direct physical interpretation, and therefore cannot be measured directly.

In this paper, we develop a model specifically for ligaments and tendons, but which will potentially be adaptable to other fibrous soft tissues. All parameters in this model *can be directly measured via experiments*. We neglect viscoelastic effects such as strain-rate dependence and hysteresis and therefore expect our model to be valid only in the low strain-rate regime where hysteretic effects are minimised. The strain energy function derived here, however, can be used to model any elastic deformation, and could potentially be incorporated into finite-strain viscoelastic models in the future. In its current form, it can easily be utilised in any finite element software that implements transversely isotropic hyperelasticity simply by replacing the existing strain energy function with that derived here.

In Section 2, we follow the method outlined by Kastelic et al. (1980) to derive stress–strain relationships for fascicles with differing fibril crimp angle distributions. By correcting the aforementioned mistake in that paper and assuming that the crimp angle distribution has a different functional form to that utilised by Kasetlic et al., we derive a simple stress–strain relationship that is used to determine the required form of the new SEF in Section 3. In Section 4, the ability of the new SEF to match experimental data is compared with that of the HGO model, and we conclude in Section 5.

## 2. The effect of fibril crimp angle distribution

Kastelic et al. (1980) derived the stress–strain response of a fascicle under uniaxial tension as a function of the microstructural arrangement of its fibrils. Here, we summarise the procedure they used, and in the following section derive an SEF for a material whose fibres have the same stress–strain response as the fascicles modelled here. We want our SEF to have the simplest possible form that still incorporates microstructural information about the fascicle, so we make some simplifications to the method described by Kastelic et al. along the way.

Kastelic et al. (1978) observed that fibril crimp angle varies throughout the radius of a fascicle, with the minimum crimp angle occurring at the centre of the fascicle, and the maximum crimp angle occurring at the fascicle's edge (see Fig. 2). If we assume that only fully extended fibrils contribute to the resistance of a fascicle to an applied load, then the resistance will increase as more fibrils become fully extended. The fibrils at the centre of the fascicle become taut first, then as the fascicle is extended, more and more fascicles within a circle of increasing radius contribute to its stiffness until the outer fibrils are finally taut. At this point, the "toe region" of the stress–strain curve ends, and the linear region begins.

Let us consider a fascicle with radius $\tilde{a}$ and define a non-dimensional radial variable $\rho = \tilde{\rho}/\tilde{a}$, where $\tilde{\rho}$ is the (dimensional) radial variable, so that $0 \le \rho \le 1$. Kastelic et al. (1980) assumed a crimp angle distribution of the form:

$$\theta_p(\rho) = \theta_o(\alpha + (1-\alpha)\rho^p), \quad (11)$$

where $\theta_p(\rho)$ is the crimp angle of the fibrils at radius $\rho$ inside the fascicle, $\theta_o$ is the crimp angle of the outermost fibrils, $\alpha$ and $p$ are angle distribution parameters, and we note that $\theta_o\alpha = \theta_i$ is the crimp angle of the fibril at the centre of the fascicle ($\rho=0$).

Assuming that the stiffness of any extra-fibrillar matrix within the fascicle is negligible, the fascicle will be stress-free until at least one of its fibrils becomes taut. We will take this configuration as our reference configuration and hence choose $\theta_i = \alpha = 0$ to obtain

$$\theta_p(\rho) = \theta_o\rho^p, \quad (12)$$

so that the fascicle appears as in Fig. 3.

The strain in the fascicle direction, as a fibril at radius $\rho$ becomes fully taut, is given by

$$\epsilon_p(\rho) = \frac{1}{\cos(\theta_p(\rho))} - 1 = \frac{1}{\cos(\theta_o\rho^p)} - 1, \quad (13)$$

and the strain required to just straighten the outer fibrils is

$$\epsilon^* = \frac{1}{\cos\theta_o} - 1. \quad (14)$$

Note that Kastelic et al. (1980) included a "crimp blunting factor", $b$, in their model, which resulted in the above expressions for $\epsilon_p(\rho)$ and $\epsilon^*$ being multiplied by $(1-b)$; however, since we are interested

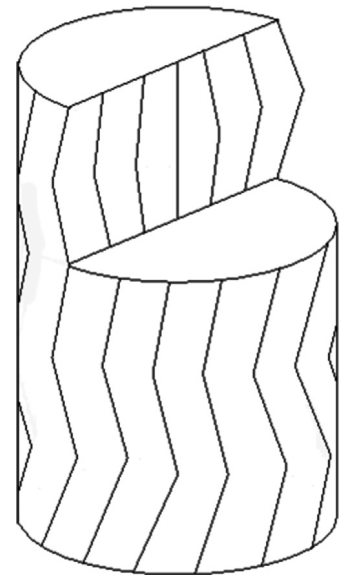

**Fig. 2.** Variation in fibril crimp angle through a fascicle (adapted from Kastelic et al., 1980).

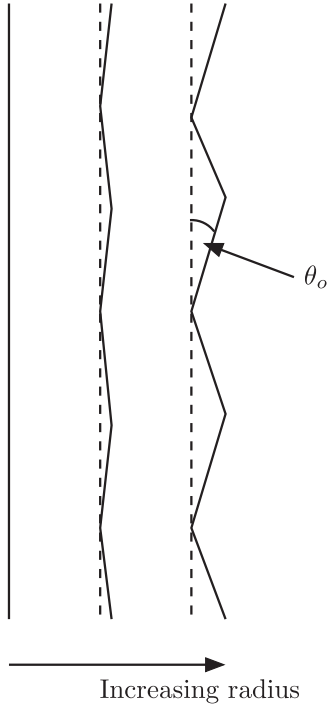

**Fig. 3.** Variation in fibril crimp angle with radius and definition of $\theta_o$.

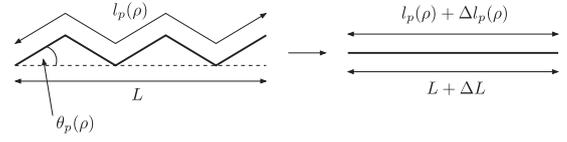

**Fig. 4.** Stretching of a fibril of initial length $l_p(\rho)$ within a fascicle of initial length $L$.

in deriving the simplest possible stress–strain relationship, we shall neglect this parameter.

At $\epsilon^*$, the non-linear "toe region" ends, and the linear elastic region begins. For a strain $\epsilon$ satisfying $0 \leq \epsilon \leq \epsilon^*$, however, there is, according to the model, an internal circular area of taut fibrils, each carrying a share of the applied load. The radius $R_p$ of this circle can be determined from (13), by equating $\epsilon = \epsilon_p(R_p)$:

$$\theta_p(R_p) = \cos^{-1}\left(\frac{1}{\epsilon+1}\right). \tag{15}$$

Upon using (12) in (15), and solving for $R_p$, we obtain

$$R_p = \left(\frac{1}{\theta_o}\cos^{-1}\left(\frac{1}{\epsilon+1}\right)\right)^{1/p}. \tag{16}$$

Fibrils outside this radius retain their crimping and experience no load. As the fascicle is stretched, this radius increases until $R_p = 1$, at which point all fibrils are finally taut.

The tensile load experienced by the fascicle $P_p$ can be determined by integration:

$$P_p = \int_0^{R_p} \sigma_p(\rho) 2\pi\rho \, d\rho, \tag{17}$$

where $\sigma_p(\rho)$ is the stress in the fibrils at radius $\rho$, and the upper limit of integration is determined by (16).

We will assume that the fibrils obey Hooke's law (as observed by Sasaki and Odajima, 1996). We note at this point that Kastelic et al. (1980) stated Hooke's law as

$$\sigma_p(\rho) = E^* \cdot \Delta\epsilon_p(\rho), \tag{18}$$

where the so-called "elastic deformation" $\Delta\epsilon_p(\rho)$ was given by

$$\Delta\epsilon_p(\rho) = \epsilon - \epsilon_p(\rho) = \epsilon - \left(\frac{1}{\cos(\theta_p(\rho))}-1\right) = \epsilon - \left(\frac{1}{\cos(\theta_o\rho^p)}-1\right). \tag{19}$$

This "elastic-deformation" is *not* the fibril strain, and as we show below, differs from the fibril strain by a quantity that is dependent on $\rho$. Therefore, $E^*$ cannot be identified as the fibril Young's modulus, and assuming that all the fibrils have the same Young's modulus, Eq. (18) cannot hold for all $\rho$. The correct form for Eq. (18) is

$$\sigma_p(\rho) = E \cdot \epsilon_p^f(\rho), \tag{20}$$

where here $\epsilon_p^f(\rho)$ is the strain in the fibrils at radius $\rho$, and now we can identify $E$ as the fibril Young's modulus. The correct expression for this strain is

$$\epsilon_p^f(\rho) = \cos(\theta_p(\rho))(\epsilon - \epsilon_p(\rho)) = (\epsilon+1)\cos(\theta_p(\rho))-1 = (\epsilon+1)\cos(\theta_o\rho^p)-1. \tag{21}$$

This can be seen by considering a fibril at radius $\rho$ of initial length $l_p(\rho)$ within a fascicle of initial length $L$ (see Fig. 4). We note that $l_p(\rho)$ and $L$ are related by

$$l_p(\rho)\cos(\theta_p(\rho)) = L. \tag{22}$$

We assume that the fascicle is stretched to a point beyond which the fibril has become taut. At this point the fascicle's length is $L + \Delta L$, the fibril's length is $l_p(\rho) + \Delta l_p(\rho)$, and

$$L + \Delta L = l_p(\rho) + \Delta l_p(\rho). \tag{23}$$

We note that the strain in the fascicle is

$$\epsilon = \frac{L+\Delta L - L}{L} = \frac{\Delta L}{L}, \tag{24}$$

and the strain in the fibril is

$$\epsilon_p^f(\rho) = \frac{l_p(\rho) + \Delta l_p(\rho) - l_p(\rho)}{l_p(\rho)} = \frac{\Delta l_p(\rho)}{l_p(\rho)}. \tag{25}$$

Therefore, using Eq. (23), we can write

$$\frac{L+\Delta L}{L} = \frac{l_p(\rho)+\Delta l_p(\rho)}{L} \tag{26}$$

$$\Rightarrow \frac{l_p(\rho)}{L}\epsilon_p^f(\rho) = \epsilon - \frac{l_p(\rho)-L}{L} = \epsilon - \epsilon_p(\rho) \tag{27}$$

$$\Rightarrow \epsilon_p^f(\rho) = \frac{L}{l_p(\rho)}(\epsilon - \epsilon_p(\rho)) = \cos(\theta_p(\rho))(\epsilon - \epsilon_p(\rho)). \tag{28}$$

Substituting (20) and (21) into (17), we can derive an expression for the (non-dimensionalised) average traction in the direction of the fascicle:

$$\frac{\tau_p}{E} = \frac{P_p}{E\pi} = 2\int_0^{R_p}((\epsilon+1)\cos(\theta_p(\rho))-1)\rho\,d\rho$$
$$= 2\int_0^{R_p}((\epsilon+1)\cos(\theta_o\rho^p)-1)\rho\,d\rho. \tag{29}$$

For certain values of $p$, (29) can be evaluated explicitly, and upon doing so for $p=1,2$, we obtain

$$\frac{\tau_1}{E} = \frac{1}{\theta_o}\left(\frac{2\sqrt{\epsilon(\epsilon+2)}}{\cos^{-1}(\epsilon+1)} - \frac{1}{(\cos^{-1}(\epsilon+1))^2} - 2\epsilon\right), \tag{30}$$

$$\frac{\tau_2}{E} = \frac{1}{\theta_o}\left(\sqrt{\epsilon(\epsilon+2)} - \frac{1}{\cos^{-1}(\epsilon+1)}\right). \tag{31}$$

In order to derive a strain energy function whose fibres have the same properties as a fascicle with the microstructure described here, we would be required to integrate the above expressions. The presence of the $\cos^{-1}(\epsilon+1)$ terms in the above expressions would make the resulting strain energy function unnecessarily complex; however, by choosing a different form for $\theta_p(\rho)$, we can obtain a simpler expression. If, instead of (12), we choose

$$\hat{\theta}_p(\rho) = \sin^{-1}(\sin(\theta_o)\rho^p), \tag{32}$$

where $\hat{\ }$ notation is used to differentiate between the new distribution and that of Kastelic et al., then, instead of (30) and (31), we obtain

$$\frac{\hat{\tau}_1}{E} = \frac{1}{3\sin^2\theta_o}\left(2\epsilon - 1 + \frac{1}{(\epsilon+1)^2}\right), \tag{33}$$

$$\frac{\hat{\tau}_2}{E} = \frac{1}{2\sin^2\theta_o}\left(\frac{\epsilon+1}{\sin^{-1}((\epsilon+1)/\sqrt{\epsilon(\epsilon+2)})} - \frac{\sqrt{\epsilon(\epsilon+2)}}{\epsilon+1}\right). \tag{34}$$

To proceed, the variation in crimp angle should be measured experimentally and plotted as a function of radius, and the parameters $p$ and $\theta_o$ should be chosen such that $\theta_p(\rho)$, or $\hat{\theta}_p(\rho)$, gives a good approximation to the observed distribution. The simple stress–strain relationship that results from using $\hat{\theta}_p(\rho)$ when $p=1$ makes this choice of distribution function preferable; however, quantitatively, there is little difference between $\theta_p(\rho)$ and $\hat{\theta}_p(\rho)$, as can be seen in Fig. 5 where we plot $\theta_p(\rho)$ and $\hat{\theta}_p(\rho)$ for $p=1,2,3$, with $\theta_o=\pi/6$. The maximum difference between the distributions is less than 0.4% of the value of $\theta_p(\rho)$ for this choice of $\theta_o$, and less than 1.8% of the value of $\theta_p(\rho)$ for any $\theta_o$ in the range $0 \leq \theta_o \leq \pi/4$.

Eqs. (30)–(34) give the contribution of the (non-dimensionalised) average traction, acting on a cross-section normal to the fascicle direction, in the direction of the fascicle, and are plotted in Fig. 6. All four expressions have the required "toe region" behaviour, and again, we see that using the new fibril distribution has a very small effect when compared with using the distribution utilised by Kastelic et al. Upon taking Taylor series expansions of these stress–strain relationships about $\epsilon=0$, we observe that there is *no linear term*. A material with the microstructure described above, therefore, should *not* be modelled as linear elastic near zero strain.

Clearly, $\hat{\tau}_1/E$ is the simplest of the stress–strain relationships above, so we use this expression to derive our SEF in the following section. We recall that (33) only holds for $0 \leq \epsilon \leq \epsilon^*$, and for $\epsilon > \epsilon^*$, we have

$$\frac{\hat{\tau}_1}{E} = \int_0^1 ((\epsilon+1)\cos(\hat{\theta}_1(\rho)) - 1)\rho\,d\rho = \beta(\epsilon+1) - 1, \quad \beta = \frac{2(1-\cos^3\theta_o)}{3\sin^2\theta_o}. \tag{35}$$

By using the relationship $\lambda = \epsilon+1$, where $\lambda$ is the stretch in the direction of the fascicle, we can write

$$\hat{\tau}_1 = \frac{E}{3\sin^2\theta_o}\left(2\lambda - 3 + \frac{1}{\lambda^2}\right), \quad 1 \leq \lambda \leq \frac{1}{\cos\theta_o}, \tag{36}$$

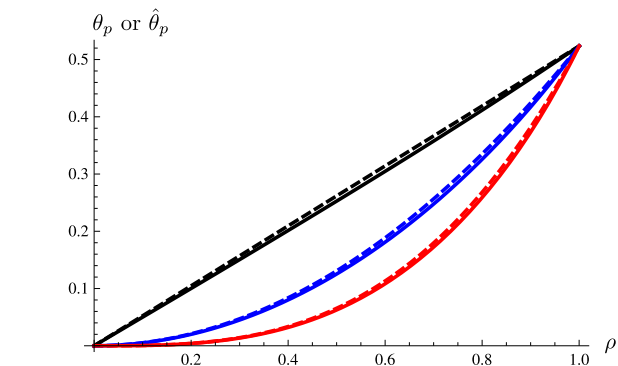

**Fig. 5.** Fibril distribution as a function of non-dimensionalised radius. Solid: $\hat{\theta}_p$, dashed: $\theta_p$. Black: $p=1$, blue: $p=2$, red: $p=3$. $\theta_o = \pi/6$.

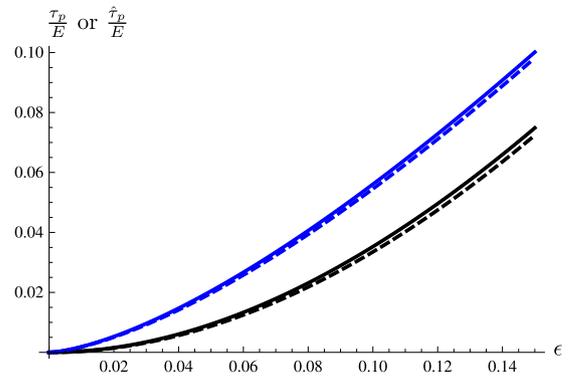

**Fig. 6.** Non-dimensionalised stress as a function of strain. Solid: $\hat{\tau}_p/E$, dashed: $\tau_p/E$. Black: $p=1$, blue: $p=2$. $\theta_o = \pi/6$.

$$\hat{\tau}_1 = E(\beta\lambda - 1), \quad \lambda > \frac{1}{\cos\theta_o}. \tag{37}$$

This form will be used in the derivation of our SEF in the following section.

## 3. Derivation of the strain energy function

In this section, we model the ligament or tendon under consideration as an incompressible, anisotropic, hyperelastic material. For a detailed account of the relevant theory the reader is referred to Holzapfel and Ogden (2010).

We characterise our material via a SEF $W$, and make the commonly used assumption that $W$ is a function of the strain invariants $I_1$ and $I_4$, only:

$$W(I_1, I_4) = (1-\phi)W_m(I_1) + \phi W_f(I_4), \tag{38}$$

where $\phi$ is the volume fraction of the fascicles. In the current context of ligaments and tendons, $W_f$ is the strain energy associated with the fascicles and $W_m$ is the strain energy associated with the matrix they run through. We note that $I_4$, which is defined in Eq. (2), can be interpreted as the square of the stretch in the fibre direction, and can be written as

$$I_4 = M_i C_{ij} M_j = M_i F_{ki} F_{kj} M_j = (\mathbf{FM})\cdot(\mathbf{FM}) = \mathbf{m}\cdot\mathbf{m} = |\mathbf{m}|^2, \tag{39}$$

where $\mathbf{m} = \mathbf{FM}$ is the push forward of $\mathbf{M}$ to the deformed configuration.

Given the SEF above, the Cauchy stress tensor takes the following form (this can be seen by taking $\partial W/\partial I_2 = \partial W/\partial I_5 = 0$ in Eq. (2.6) of Holzapfel and Ogden, 2010):

$$\mathbf{T} = -Q\mathbf{I} + 2\frac{\partial W}{\partial I_1}\mathbf{B} + 2\frac{\partial W}{\partial I_4}\mathbf{m}\otimes\mathbf{m}, \tag{40}$$

where $Q$ is a Lagrange multiplier associated with the incompressibility constraint, $\mathbf{I}$ is the identity tensor, and $\mathbf{B} = \mathbf{FF}^T$ is the left Cauchy–Green tensor. In the above, $\mathbf{T}_f = 2W_4\mathbf{m}\otimes\mathbf{m}/\phi$ is the component of the Cauchy stress associated with the fascicles, which will be used to derive an expression equivalent to $\hat{\tau}_1$ in (36) and (37).

The traction associated with $\mathbf{T}_f$ acting on a face normal to the direction of the fascicles in the deformed configuration (i.e. a face with unit normal given by $\hat{\mathbf{m}} = \mathbf{m}/|\mathbf{m}|$) is

$$\mathbf{t}_f = \mathbf{T}_f\cdot\left(\frac{\mathbf{m}}{|\mathbf{m}|}\right) = \frac{2}{\phi}\frac{\partial W}{\partial I_4}(\mathbf{m}\otimes\mathbf{m})\cdot\left(\frac{\mathbf{m}}{|\mathbf{m}|}\right)$$

$$= \frac{2}{\phi}\frac{\partial W}{\partial I_4}\mathbf{m}\frac{\mathbf{m}\cdot\mathbf{m}}{|\mathbf{m}|} = \frac{2}{\phi}\frac{\partial W}{\partial I_4}|\mathbf{m}|\mathbf{m}, \tag{41}$$

and the component of this traction in the direction of the fascicles is given by

$$\frac{\mathbf{t}_f \cdot \mathbf{m}}{|\mathbf{m}|} = \left(\frac{2}{\phi}\frac{\partial W}{\partial I_4}|\mathbf{m}|\mathbf{m}\right) \times \left(\frac{\mathbf{m}}{|\mathbf{m}|}\right) = \frac{2}{\phi}\frac{\partial W}{\partial I_4}|\mathbf{m}|^2 = \frac{2}{\phi}\frac{\partial W}{\partial I_4}I_4 = 2I_4\frac{dW_f}{dI_4}. \quad (42)$$

By equating (42) with (36) and (37), we therefore obtain two equations for the required form of $W_f$:

$$2I_4\frac{dW_f}{dI_4} = \frac{E}{3\sin^2\theta_o}\left(2\lambda - 3 + \frac{1}{\lambda^2}\right), \quad 1 \leq \lambda \leq \frac{1}{\cos\theta_o}, \quad (43)$$

$$2I_4\frac{dW_f}{dI_4} = E(\beta\lambda - 1), \quad \lambda > \frac{1}{\cos\theta_o}. \quad (44)$$

Since $I_4$ is the square of the stretch in the fibre direction, $\lambda = \sqrt{I_4}$, so that

$$2I_4\frac{dW_f}{dI_4} = \frac{E}{3\sin^2\theta_o}\left(2\sqrt{I_4} - 3 + \frac{1}{I_4}\right), \quad 0 \leq I_4 \leq \frac{1}{\cos^2\theta_o}, \quad (45)$$

$$2I_4\frac{dW_f}{dI_4} = E(\beta\sqrt{I_4} - 1), \quad I_4 > \frac{1}{\cos^2\theta_o}, \quad (46)$$

and, hence

$$\frac{dW_f}{dI_4} = \frac{E}{6\sin^2\theta_o}\left(\frac{2}{\sqrt{I_4}} - \frac{3}{I_4} + \frac{1}{I_4^2}\right), \quad 1 \leq I_4 \leq \frac{1}{\cos^2\theta_o}, \quad (47)$$

$$\frac{dW_f}{dI_4} = \frac{E}{2}\left(\beta\frac{1}{\sqrt{I_4}} - \frac{1}{I_4}\right), \quad I_4 > \frac{1}{\cos^2\theta_o}. \quad (48)$$

Integrating and (47) and (48) with respect to $I_4$, we obtain the required form for the anisotropic component of our SEF:

$$W_f = \frac{E}{6\sin^2\theta_o}\left(4\sqrt{I_4} - 3\log(I_4) - \frac{1}{I_4} + \gamma\right), \quad 1 \leq I_4 \leq \frac{1}{\cos^2\theta_o}, \quad (49)$$

$$W_f = E\left(\beta\sqrt{I_4} - \frac{1}{2}\log(I_4) + \eta\right), \quad I_4 > \frac{1}{\cos^2\theta_o}, \quad (50)$$

where $\gamma$ and $\eta$ are constants of integration, which must be chosen to satisfy $W_f|_{I_4=1} = 0$ and to ensure that $W_f$ is continuous at $I_4 = 1/\cos^2\theta_o$. Upon applying these conditions, we obtain

$$\gamma = -3, \quad \eta = -\frac{1}{2} - \frac{\cos^2\theta_o}{\sin^2\theta_o}\log\left(\frac{1}{\cos\theta_o}\right). \quad (51)$$

Finally, we note that for $I_4 < 1$, $W_f = 0$. Therefore, we now have an explicit expression for the anisotropic part of our SEF.

To determine the form of the isotropic component of our SEF, we follow the convention used by Holzapfel et al. (2000). In the modelling of arteries, a neo-Hookean component of the SEF is commonly used to model the arterial elastin which makes up the matrix and this approach is backed up by the work of Gundiah et al. (2007). We make the assumption that the loose connective tissue that forms the matrix in ligaments and tendons has similar mechanical properties to arterial elastin so that the neo-Hookean model is still reasonable. We, therefore, have

$$W_m(I_1) = \frac{\mu}{2}(I_1 - 3), \quad (52)$$

where $\mu > 0$ is a stress-like parameter, which for a neo-Hookean material in isolation may be identified as the ground state shear modulus.

We have now determined the complete form of our SEF, which is written explicitly as follows:

$$W = (1-\phi)\frac{\mu}{2}(I_1 - 3), \quad I_4 < 1, \quad (53)$$

$$W = (1-\phi)\frac{\mu}{2}(I_1 - 3) + \frac{\phi E}{6\sin^2\theta_o}\left(4\sqrt{I_4} - 3\log(I_4) - \frac{1}{I_4} - 3\right), \quad 1 \leq I_4 \leq \frac{1}{\cos^2\theta_o}, \quad (54)$$

$$W = (1-\phi)\frac{\mu}{2}(I_1 - 3) + \phi E\left(\beta\sqrt{I_4} - \frac{1}{2}\log(I_4) + \eta\right), \quad I_4 > \frac{1}{\cos^2\theta_o}, \quad (55)$$

where $\beta$ is defined in (35), and $\eta$ is defined in (51). We note that all of the parameters in the above model can be measured. The outer fibril crimp angle of a fascicle $\theta_o$ can be measured via scanning electron microscopy (Yahia and Drouin, 1989), the fibril Young's modulus $E$ can be measured via X-ray diffraction (Sasaki and Odajima, 1996), or atomic force microscopy (Svensson et al., 2012), and it should soon be possible to measure the fibre volume fraction $\phi$ and fibre alignment vector $\mathbf{M}$ via X-ray computed tomography (Shearer et al., 2014). The only parameter which is not straightforward to determine is $\mu$; however, the techniques used by Gundiah et al. (2007) to determine the value of this parameter for arterial elastin could potentially be adapted to do the same for ligament or tendon matrix.

Finally, we note that the SEF above behaves isotropically in the small strain limit. This may seem unusual for a transversely isotropic SEF, but this behaviour arises due to the microstructure of the fascicles, which do not have a linear term in their stress–strain relationship for small strains, as discussed in Section 2.

## 4. Comparison with the HGO model

In this section, we compare the new model (53)–(55) with the HGO model (1). The two models are plotted against stress–strain data for human patellar tendon taken from Johnson et al. (1994). Johnson et al. plotted separate results for samples taken from younger (aged 29–50) and older (aged 64–93) donors. We fit the models to the experiments performed on the younger samples. The patellar tendon was chosen due to the fact that its fascicles are very strongly aligned with its longitudinal axis (Shearer et al., 2014), so that the fibre alignment vector $\mathbf{M}$ could be chosen to coincide with the tendon's axis in the model. In this simple example, we will model the patellar tendon as a circular cylinder. This is not the case in reality, but for the purpose of illustrating the model developed above, this geometry will be sufficient.

Consider a cylinder with radius $A$ and length $B$ in the undeformed configuration. We describe its geometry in this configuration in terms of the polar coordinates $(R, \Theta, Z)$, so that $0 \leq R \leq A$, $0 \leq \Theta \leq 2\pi$, $0 \leq Z \leq B$. After the application of a homogeneous longitudinal stretch, its geometry in the deformed configuration is described by the polar coordinates $(r, \theta, z)$, so we have $0 \leq r \leq a$, $0 \leq \theta \leq 2\pi$, $0 \leq z \leq b$, where $a$ and $b$ are the deformed counterparts of $A$ and $B$, respectively. The deformation, then, is described by

$$r = \frac{R}{\sqrt{\zeta}}, \quad \theta = \Theta \quad \text{and} \quad z = \zeta Z, \quad (56)$$

where $\zeta = b/B$ is the uniform stretch in the longitudinal direction and the first equation is a consequence of the fact that we are using incompressible constitutive models. Due to the symmetry of the imposed deformation, the deformation gradient tensor will be diagonal and its entries will be the principal stretches:

$$\mathbf{F} = F_{ij}\mathbf{e}_i \otimes \mathbf{E}_J, \quad F_{ij} = \begin{pmatrix} \zeta^{-1/2} & 0 & 0 \\ 0 & \zeta^{-1/2} & 0 \\ 0 & 0 & \zeta \end{pmatrix}, \quad (57)$$

where $\mathbf{e}_i$, $i = (r, \theta, z)$, and $\mathbf{E}_J$, $J = (R, \Theta, Z)$, are the deformed and undeformed unit vectors in the radial, azimuthal and longitudinal

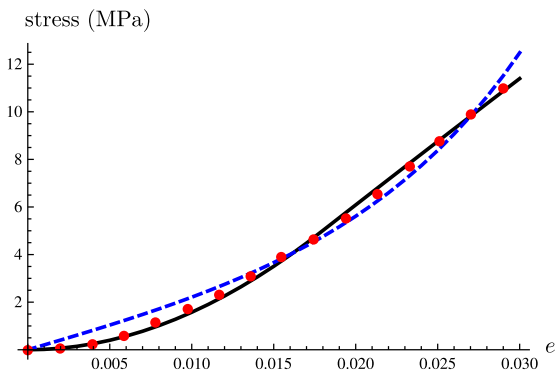

**Fig. 7.** Stress–strain curves comparing the ability of the new model to reproduce experimental stress–strain data for human patellar tendon taken from Johnson et al. (1994) with that of the HGO model. Solid black: new model, dashed blue: HGO model, red circles: experimental data. Parameter values: $c = (1-\phi)\mu = 0.01$ MPa, $k_1 = 25$ MPa, $k_2 = 183$, $\phi E = 552$ MPa, $\theta_o = 0.19$ rad $= 10.7°$.

directions, respectively, and the left Cauchy–Green tensor is given by

$$\mathbf{B} = B_{ij}\mathbf{e}_i \otimes \mathbf{e}_j, \quad B_{ij} = \begin{pmatrix} \zeta^{-1} & 0 & 0 \\ 0 & \zeta^{-1} & 0 \\ 0 & 0 & \zeta^2 \end{pmatrix}. \quad (58)$$

Since the fascicles of the patellar tendon are aligned with its longitudinal axis, we have $\mathbf{M} = \mathbf{E}_Z$, and, therefore, $\mathbf{m} = \zeta\mathbf{e}_z$. The strain invariants $I_1$ and $I_4$ can also be calculated:

$$I_1 = \frac{2}{\zeta} + \zeta^2, \quad I_4 = \zeta^2. \quad (59)$$

Upon using Eqs. (57)–(59) in Eq. (40), solving the static equilibrium equations, div $\mathbf{T} = \mathbf{0}$, and applying traction-free boundary conditions on $r = a$, we obtain explicit expressions for the longitudinal stresses $T_{zz}^{HGO}$ and $T_{zz}$:

$$T_{zz}^{HGO} = \begin{cases} c(\zeta^2 - \zeta^{-1}), & \zeta < 1, \\ c(\zeta^2 - \zeta^{-1}) + 4k_1\zeta^2(\zeta^2-1)e^{k_2(\zeta^2-1)^2}, & \zeta \geq 1, \end{cases} \quad (60)$$

$$T_{zz} = \begin{cases} (1-\phi)\mu(\zeta^2 - \zeta^{-1}), & \zeta < 1, \\ (1-\phi)\mu(\zeta^2 - \zeta^{-1}) + \dfrac{\phi E}{3\sin^2\theta_o}(2\zeta - 3 + \zeta^{-2}), & 1 \leq \zeta \leq \dfrac{1}{\cos\theta_o}, \\ (1-\phi)\mu(\zeta^2 - \zeta^{-1}) + \phi E(\beta\zeta - 1), & \zeta > \dfrac{1}{\cos\theta_o}, \end{cases}$$
$$(61)$$

where the superscript HGO is to differentiate the HGO model from that presented above. In Fig. 7, we plot the nominal stresses $S_{zz}^{HGO} = T_{zz}^{HGO}/\zeta$ and $S_{zz} = T_{zz}/\zeta$ (which give the force per unit *undeformed* area, to match the experimental data) as a function of the engineering strain $e = \zeta - 1$, along with the patellar tendon stress–strain data taken from Fig. 4 of Johnson et al. (1994). We note that this experimental data clearly displays non-linear behaviour, even for strains of less than 1%.

Since the stiffness of ligament and tendon matrix is insignificant compared with that of its fascicles, $c$ and $(1-\phi)\mu$ were chosen to be small ($c = (1-\phi)\mu = 0.01$ MPa), so that the contribution of the matrix material to the overall stress was negligible. The values of $k_1$, $k_2$, $\phi E$ and $\theta_o$ were fitted to the experimental data using the FindFit function in Mathematica 7 (Wolfram Research, Inc., Champaign, Illinois, 2008) subject to the conditions: $k_1 > 0$, $\phi E > 0$, $0 \leq \theta_o \leq \pi/2$. The values determined were $k_1 = 25$ MPa, $k_2 = 183$, $\phi E = 558$ MPa, and $\theta_o = 0.19$ rad $= 10.7°$. We quantify the effectiveness of the two models in terms of the relative error, defined at a given strain by

$$\delta = |S_{zz}^{exp} - S_{zz}|/|S_{zz}^{exp}|, \quad \delta^{HGO} = |S_{zz}^{exp} - S_{zz}^{HGO}|/|S_{zz}^{exp}|, \quad (62)$$

where $S_{zz}^{exp}$ is the experimental stress, and the absolute error, defined at a given strain by

$$\Delta = |S_{zz}^{exp} - S_{zz}|, \quad \Delta^{HGO} = |S_{zz}^{exp} - S_{zz}^{HGO}|. \quad (63)$$

The average relative error for the new model was $\overline{\delta} = 0.053$, and that of the HGO model was $\overline{\delta}^{HGO} = 0.57$, and the average absolute error for the new model was $\overline{\Delta} = 0.12$ MPa, and that of the HGO model was $\overline{\Delta}^{HGO} = 0.31$ MPa. Hence, the relative error associated with the new model is less than 10% of that of the HGO model and the absolute error associated with the new model is less than 41% of that of the HGO model.

## 5. Discussion

In this paper, we derive a new SEF for the hyperelastic modelling of ligaments and tendons. The form of this SEF is determined from the stress–strain response of fascicles as a function of their microstructure, using the method described by Kastelic et al. (1980). In Section 2, a simple, explicit form for this stress–strain behaviour is determined by altering the form of the fibril crimp angle distribution assumed by Kastelic et al. and the corresponding SEF is derived in Section 3. Crucially, all of the parameters in this new model can potentially be measured experimentally.

In Section 4, we compare the ability of the new model to reproduce experimental stress–strain data for human patellar tendon taken from Johnson et al. (1994) with that of the HGO model proposed in Holzapfel et al. (2000). Since independent measurements of the collagen volume fraction $\phi$, fibril stiffness $E$, and average outer fibril crimp angle $\theta_o$ are not readily available, these parameters are chosen to fit the experimental data. The new model is found to be more successful at modelling this experimental data, and the values of the parameters used in the new model are physically realistic. The value of $\phi E$ used above gives a value of $E$ that lies well within the range of reported values of 0.07 (Yang et al., 2008)–5.1 GPa (Cusack and Miller, 1979), assuming $0.11 \leq \phi \leq 1$. To the author's knowledge, data on typical values of $\theta_o$ is not readily available; however, the predicted value of $\theta_o = 10.7°$ seems reasonable. We note that the new model performs particularly well in terms of relative error. This is due to the better performance of the new model within the "toe region" of the stress–strain curve when compared with the HGO model.

In order to fully validate this model, it will be important to independently measure $\phi$, $E$ and $\theta_o$ for a given ligament or tendon, input these quantities into the model and compare the predicted results with experimental stress–strain data for that ligament or tendon.

## Conflict of interest statement

The author has no conflicts of interest to declare.

## Acknowledgments

The author would like to thank EPSRC for funding this work through Grant EP/L017997/1, and Dr W.J. Parnell and Dr A.L. Hazel for their feedback and several constructive discussions.